\documentclass[aip,notitlepage, floatfix, jcp, twocolumn, reprint]{revtex4-1}
\usepackage{amssymb}
\usepackage{amsfonts}
\usepackage{amsmath}
\usepackage{graphicx}
\usepackage{enumerate}
\usepackage[usenames,dvipsnames]{color}
\usepackage{ulem}

\begin{document}

\title{Nucleation of colloids and macromolecules: does the nucleation
pathway matter?}
\author{James F. Lutsko}
\affiliation{Center for Nonlinear Phenomena and Complex Systems, Code Postal 231,
Universit\'{e} Libre de Bruxelles, Blvd. du Triomphe, 1050 Brussels, Belgium}
\email{jlutsko@ulb.ac.be}
\homepage{http://www.lutsko.com}

\begin{abstract}
A recent description of diffusion-limited nucleation based on fluctuating hydrodynamics that extends classical nucleation theory predicts a very non-classical two-step scenario whereby nucleation is most likely to occur in spatially-extended, low-amplitude density fluctuations. In this paper, it is shown how the formalism can be used to determine the maximum probability of observing \emph{any} proposed nucleation pathway, thus allowing one to address the question as to their relative likelihood, including of the newly proposed pathway compared to classical scenarios. Calculations are presented for the nucleation of high-concentration bubbles in a low-concentration solution of globular proteins and it is found that the relative probabilities (new theory compared to classical result) for reaching a critical nucleus containing $N_c$ molecules scales as $e^{-N_c/3}$ thus indicating that for all but the smallest nuclei, the classical scenario is extremely unlikely.
\end{abstract}

\date{\today }
\maketitle

\section{Introduction}

Nucleation - whether homogeneous or heterogeneous - is a paradigmatic
example of self assembly. It occurs when a physical system can be in two or
more states that are separated by free-energy barriers. If the system is
initially in one of the metastable states, thermal fluctuations can drive it
over the free energy barrier and into a more stable state. Because the
energy needed to overcome the barrier scales with the spatial size of the
system, the process occurs locally via the formation of a finite sized
cluster or nucleus. In many processes of interest such as the
crystallization of proteins from solution\cite%
{GuntonBook,VekilovCGDReview2004}, the formation of snowflakes\cite{Libbrecht}  and the crystallization of polymorphic
solids\cite{Polymorph} the system may pass through one or  more intermediate
metastable states before arriving at the final, stable state. In fact, the
heuristic known as Ostwald's rule of stages specifically states that a
system will pass in turn from one state to another having the next lowest
free energy until it reaches the minimal energy state. Various arguments can
be given in support of this rule\cite{Ostwald} and they can  be grouped into
two classes: either it arises because the barriers separated ``similar''
states  are smaller than those separating disparate states or it arises due
to the kinetics of the transition. The former reason is an application of
another heuristic known as the Stranski-Totomanow conjecture which states
that the observed transition will be the one corresponding to the minimal
energy barrier\cite{Stranski}. The latter is more difficult to characterize
as it may depend on microscopic details of the various states and the
dynamics of the system. Recent work on model systems indicates that,
depending on the free energy landscape and the dynamics, either Ostwald or
Stranski-Totomanow or both may be correct or not\cite{Whitelam}.

The common issue at question is the description of the \emph{nucleation
pathway}.  Clearly, a theoretical description for the nucleation pathway
that goes beyond the empirical heuristics mentioned above must be based on a
dynamical description of the nucleation including the role of thermal
fluctuations, mass and energy transport and the structure of the various
phases. A framework for such a dynamical formulation of nucleation has
recently been described\cite{Lutsko_JCP_2011_Com,Lutsko_JCP_2012}. For large clusters, it was
shown to reproduce classical nucleation theory (CNT) in the weak-noise limit. For a diffusion-limited dynamics - appropriate for
the description of colloids and macromolecules in solution - it was shown
that the relative probability of different nucleation pathways could be
easily calculated and that the most likely path (MLP) could be determined by
steepest-descent on the free energy surface. This framework can be
contrasted with numerous proposals for determining the nucleation pathway
based solely on the properties of the free energy surface with no dynamical
input, see e.g. \onlinecite{talanquer:5190, LutskoBubble1,LutskoBubble2,
Ghosh, Corti_PRL, Philippe, EvansArcherNucleation}. As discussed below,
because of the
heuristic nature of the latter, evaluating their relative merits has proven difficult.

The purpose of this paper is to illustrate the comparison of different
candidate nucleation pathways by means of their relative probability. The
particular example investigated is the formation of high concentration
droplets in a low-concentration solution of globular proteins. This process
is analogous to the vapor-liquid transition in simple fluids and, while
being of intrinsic interest as part of the process of crystallization in
globular proteins, has the practical advantage of allowing for a relatively
simple theoretical description as described in Ref. %
\onlinecite{Lutsko_JCP_2011_Com,Lutsko_JCP_2012} and below. For this
problem, the recently developed dynamical theory of nucleation predicts a
very different nucleation pathway than do the older DFT-based theories. The
new prediction is that the process of droplet nucleation involves two steps:
first, a long-wavelength density fluctuation forms and then a nucleation
event takes place within this fluctuation. The older, classical view is that
nucleation begins with a spatially localized cluster that grows
monotonically. From a point of view focused on the comparison of the
theories, it is therefore of interest to compare the relative likelihood of
these two pathways so as to determine whether the difference between them is
qualitative or quantitative. More broadly, in the context of the general
theory of nucleation this allows us to address the question of how important
it is to choose the ``right'' path when there are multiple possible paths
that pass through the (unique) critical cluster. Note that a feature of the
heuristic rules is that they are phrased entirely in terms of the free
energy maxima (the barriers) and  minima (the states) and, hence, the
provide no guidance on this question. In this sense, they are in accord with
the common intuition that all that matters in nucleation are the free-energy
extrema which, in addition to their use in choosing  pathways, are also the
only relevant quantities entering the Classical Nucleation Theory (CNT) for
nucleation rates\cite{Kashchiev}. Here, it will be shown that even paths
beginning and ending on the same states and passing over the same barriers
can have wildly different probabilities of occurrence. The paths compared
will be the most likely path as determined from the dynamical theory and a
path determined from one of the non-dynamical, free-energy methods which, by
means of direct comparison to simulation\cite{LutskoBubble2}, is known to
give a very accurate description of the free energy barrier for nucleation
in proteins. Since this method has also recently been applied to the study
of nucleation of wetting films on curved substrates\cite{WettingCurved} and
of ordered phases of block copolymers\cite{copolymers}, where multiple
candidate pathways were found, it is of particular interest to use as a
test-case. This paper therefore serves two puposes:  (i) to introduce a method of comparing the likelihood of candidate nucleation pathways, however they are arrived at; and (ii)to use this method to determine whether the proposed non-classical pathway is significantly different (more probable) than a ``classical'' alternative.

In Section II, the elements of the theoretical description are reviewed.
Section III describes the application to the nucleation of protein-rich
droplets in solution, including a detailed comparison of the different
candidate pathways and a computation of their relative probabilities. The
paper ends with a brief discussion of our conclusions.

\section{Theory}

The present theoretical development is concerns a collection of particles - molecules or colloidal particles -
that interact with one another via a prescribed pair potential and which are also subject to random, Brownian forces. This
is a simple model for large particles in a bath of small particles wherein the effect of the bath or solution is incorporated
via the effective interaction between the large particles and random (Brownian) forces acting on the large particles. As such, the bath
is not explicitly represented except through the amplitude of the random noise which in turn determines the (low-concentration) diffusion constant for the
large particles. For this reason, the concentration of the large particles is equivalent to their density and the two terms will be used interchangeable in the following. Further details of the microscopic model can be found in Ref. \onlinecite{Lutsko_JCP_2012}. The  fundamental quantity with which the phase transition is characterized is then
the local concentration (or number density), $\rho\left( \mathbf{r}\right) $.  The density/concentration is commonly
assumed to be spherically symmetric and this assumption will be used
throughout the present development. The initial, metastable, system is
characterized by a uniform density $\rho\left( r\right) =\rho_{i}$ where $%
\rho_{i}$ is a minimum of the bulk free energy. The new phase is also
characterized by a uniform density, $\rho\left( r\right) =\rho_{f}$, where $%
\rho_{f}$ is the global minimum of the bulk free energy. We assume
throughout the existence of a Helmholtz free energy functional, $F\left[ \rho%
\right] $, so that, e.g., in the grand canonical ensemble the appropriate
free energy is $\Omega\left[ \rho\right] =F\left[ \rho\right] -\mu N$ where $%
N=\int\rho\left( \mathbf{r}\right) d\mathbf{r}$ is the total number of
particles. For a uniform density the free energy becomes an ordinary
function, $F\left[ \rho\left( \mathbf{r}\right) =\rho_{i}\right] \equiv
F\left( \rho _{i}\right) $, and the conditions for the uniform phases to be
minima are the usual relations $\frac{1}{V}F^{\prime}\left( \rho_{i}\right)
=\mu=\frac {1}{V}F^{\prime}\left( \rho_{f}\right) $.

Nucleation proceeds by the formation of a cluster consisting of the new
phase which grows until it consumes the entire system. Despite the fact that
the new phase is energetically favored, small clusters are unstable due to
the dominance of surface tension effects. When a cluster is sufficiently
large, the lowering of the cluster energy due to increasing the size of the
bulk region inside the cluster outweighs the cost of increasing the surface
area and growth is favored. These regimes are separated by a saddle point in
the free energy called the critical cluster which necessarily satisfies the
relation%
\begin{equation}
\frac{\delta\Omega\left[ \rho\right] }{\delta\rho\left( \mathbf{r}\right) }%
=0.
\end{equation}
There are generally two approaches to representing the density function. One
is to simply discretize space by introducing a lattice of points $%
r_{i}=i\Delta$ so that one works with a series of values $\rho_{i}\equiv
\rho\left( i\Delta\right) $. An alternative is to use parametrized
functional form such as a hyperbolic tangent,%
\begin{equation}
\rho\left( r\right) =\rho_{0}+\frac{\rho_{\infty}-\rho_{0}}{1+\exp\left( 
\frac{R-r}{w}\right) }
\end{equation}
or an exponential form%
\begin{eqnarray}
\rho\left( r\right) &=& \left[ \rho_{0}-\frac{\rho_{0}-\rho_{\infty}}{2}%
\exp\left( \frac{r-R}{w}\right) \right] \Theta\left( R-r\right) \\
&&+\left[\rho_{\infty} +\frac{\rho_{0}-\rho_{\infty}}{2}\exp\left( \frac{R-r%
}{w}\right) \right] \Theta\left( r-R\right)  \notag
\end{eqnarray}
where, in both cases, there are four parameters: a radius, $R$, a width, $w$%
, the interior density, $\rho_{0}$ that characterizes the central density
for large $(R\gg w)$ clusters and the density far from the cluster, $%
\rho_{\infty }$. It is easy to show that in the thermodynamic limit the
latter must be a minimum of the free energy so that we will normally have
that it is equal to the initial density $\rho_{\infty}=\rho_{i}$. The
advantage of the first method, discretization, is that it is clear that one
can approach the continuum limit by decreasing the lattice spacing whereas
the advantage of the second method, parametrization, is that one can hope
to get good results with relatively few parameters. In fact, the two methods
can be viewed as two different approaches to parametrization and other
possibilities - such as representation in terms of Fourier components or
projection onto some other set of basis functions - have the same
characteristic. We can therefore without loss of generality assume that the
density field is represented by a collection of $N$ parameters denoted
generically as $x_{i}$ so that $\rho\left( r\right) =f\left( r;\mathbf{x}%
\right) $, for some function $f$. The change in the density profile as a
function of time therefore becomes a change in the parameters so that more
generally we have $\rho\left( r;t\right) =f\left( r;\mathbf{x}\left(
t\right) \right) $. Thus, specification of the evolution of the parameters, $%
\mathbf{x}\left( t\right) $, corresponds to the prescription of a path in
density space. The same notion holds when the path is parametrized by some
other quantity rather than time - e.g. the equivalent of a reaction
coordinate.

To give a concrete illustration, Ghosh and Ghosh use the exponential profile
with the radius acting as the reaction coordinate. They determine the other
parameters by minimizing the free energy while holding the radius constant
thus parametrize the path by the radius. However, this is not a unique
prescription since one could equally well parametrize by the excess number
of particles in the cluster, 
\begin{equation}
\Delta N\equiv\int\left( \rho\left( r\right) -\rho_{\infty}\right) d\mathbf{r%
}
\end{equation}
and, as pointed out previously, these need not be the same since one could
increase $\Delta N$ while holding the radius constant and increasing the
width. This ambiguity is the fundamental problem with methods based solely
on free energy considerations:\ there is no obvious method to determine
which approach is preferable (or more pertinently, which is chosen by
Nature).

Previously, it was shown that under the assumptions listed above, the MLP
can be determined by gradient descent on the free energy surface. This means
that first the saddle point is located and then one solves%
\begin{equation}
\frac{dx_{i}}{dt}=\pm g_{ij}^{-1}\left( \mathbf{x}\right) \frac {\partial\Omega}{\partial x_{j}} \label{gradient}
\end{equation}
where the sign is chosen according to whether one is moving uphill from the
initial state, $\rho_{i}$, to the saddle point (plus sign) or downhill from
the saddle point to the final state, (minus sign). Alternatively, one can
start at the saddle point and solve this equation with the minus sign to
determine the two halves of the path. The information about dynamics is
contained in the matrix of kinetic coefficients, $g_{ij}^{-1}$, which are calculated as the inverse of
\begin{equation}
g_{ij}\left( \mathbf{x}\right) =\int_{0}^{\infty}\frac{1}{4 \i r^2\rho\left( r;%
\mathbf{x}\right) }\frac{\partial m\left( r;\mathbf{x}\right) }{\partial
x_{i}}\frac{\partial m\left( r;\mathbf{x}\right) }{\partial x_{j}}dr
\end{equation}
where the cumulative mass is%
\begin{equation}
m\left( r;\mathbf{x}\right) =4\pi\int_{0}^{r}\rho\left( r^{\prime}\right)
r^{\prime2}dr^{\prime}.
\end{equation}
Another interpretation of Eq.(\ref{gradient}) is that the MLP is determined by steepest descent on the free energy surface with the matrix
$g_{ij}$ playing the role of a Riemannian metric. 
In general, the probability density for any path, $\rho\left( r;\mathbf{x}%
\left( t\right) \right) $for $0\leq t\leq T$, is given by%
\begin{equation}
P\left[ \mathbf{x}\right] =\mathcal{N}\exp\left( -\frac{1}{2}S\left[\mathbf{x}\right]\right) \label{path1}
\end{equation}
where the action is 
\begin{equation}
S\left[ \mathbf{x}\right] =\int_{0}^{T} \mathcal{L}(\mathbf{x},\mathbf{\dot{x}}) dt,
\end{equation}
the Lagrangian is 
\begin{equation}
\mathcal{L}(\mathbf{x},\mathbf{\dot{x}})  =\frac{1}{2}\left( \frac{dx_{i}}{dt}-g_{ij}^{-1}\frac{\partial\Omega}{%
\partial x_{j}}\right) g_{il}\left( \frac{dx_{l}}{dt}-g_{lk}^{-1}\frac{%
\partial\Omega }{\partial x_{k}}\right)
\end{equation}
and where the normalization constant, $\mathcal{N}$, is independent of the path but otherwise unknown.
Given two paths, $\mathbf{x}\left( t\right) $ and $\mathbf{y}\left( t\right) 
$, their relative probability can be calculated using these expressions
provided they are parametrized in the same way. Note that Eq.(\ref{path1})
gives the \emph{probability density} and not the \emph{probability} of the
path. The latter would actually be $P\left[ \mathbf{x}\right] D\mathbf{x}$
where $D\mathbf{x}$ is the path measure. For example, if time were
discretized using $M+1$ values $t_{i}=i\left( T/M\right) $ and if the
parameters are then given by $x_{ij}\equiv x_{i}\left( t_{j}\right) $ then
the measure would be $D\mathbf{x}=\prod \limits_{i=1}^{N}\prod
\limits_{j=0}^{M}dx_{ij}$. This factor is irrelevant for computing the
relative probability of two paths described by the same parametrization
since it would drop out of the ratio leaving the ratio of the probability
densities, $P\left[ \mathbf{x}\right] /P\left[ \mathbf{y}\right] $. On the
other hand, it is clear that one cannot meaningfully compare two paths based
on \emph{different} parametrization schemes since the measure, and the
normalization factor, would in general be different.

It may still be possible to compare the paths in an approximate manner by
approximately translating one parametrization scheme into another. For
example, \ a path based on a discretization of the density on a set of $N$
lattice points, $r_{i}$, cannot be directly compared to one based on a
discretization over $2N$ lattice points, $r_{i}^{\prime}$. However, one can
translate either parametrization into the language of the other by, e.g.,
using cubic-spline interpolation over the $N$-point profiles to evaluate the
density at the lattice positions of the $2N$-point discretization.

Another question is whether there is any way to compare the probabilities of
pathways that are not determined from the dynamical model? In particular,
one would like to be able to evaluate the utility of the large number of
existing heuristic approaches for the determination of nucleation pathways
which typically give a path in terms of a reaction coordinate, which will be
called $s$, rather than the time?\ For example, if the path is determined by
minimizing the free energy at constant excess particle number, $N$, for
different values of $N$ ranging from zero to the mass of the critical
cluster, then the reaction coordinate, $s$, would be $N$. It would seem that
the only physical meaning of such a path is that a real system would follow
it in the course of time so that if the path $\mathbf{x}\left( s\right) $
goes from the initial state at $s=0$ to the saddle point at $s=s_{max}$ then, this
must mean that the physical system begins at $\rho\left( r;t=0\right)
=\rho\left( r;\mathbf{x}\left( s=0\right) \right) $ and ends at the saddle
point at $\rho\left( r;t=T\right) =\rho\left( r;\mathbf{x}\left( s=s_{max}\right)
\right) $. Since both time and the reaction coordinate vary monotonically
during this processes, it must be that they are related so that we can write 
\begin{equation}
\frac{d\rho}{dt}=\frac{ds}{dt}\frac{d\rho}{ds}\equiv v(s)\frac{d\rho}{ds}
\end{equation}
with $v(s)\equiv ds/dt$ being the speed along the path. Writing substituting
this into the expression for the path probability we get%
\begin{widetext}
\begin{equation}
P\left[  \rho\right]  =\mathcal{N}\exp\left(  -\frac{1}{4}\int_{0}^{s_{max}}\left(
v(s)\frac{dx_{i}}{ds}-g_{ij}^{-1}\frac{\partial\Omega}{\partial x_{j}}\right)
g_{il}\left(  v(s)\frac{dx_{l}}{ds}-g_{lk}^{-1}\frac{\partial\Omega}{\partial
x_{k}}\right)  v^{-1}(s)ds\right)
\end{equation}
\end{widetext}
We can now maximize the probability with respect to variations in $v(s)$ to
find that%
\begin{equation}
\frac{dx_{i}}{ds}g_{il}\frac{dx_{l}}{ds}-\frac{\partial\Omega}{\partial x_{j}%
}g_{kj}^{-1}\frac{\partial\Omega}{\partial x_{k}}\left( v(s)\right) ^{-2}=0
\end{equation}
or%
\begin{equation}
\sqrt{\frac{\frac{dx_{i}}{ds}g_{il}\frac{dx_{l}}{ds}}{\frac{\partial\Omega }{%
\partial x_{j}}g_{kj}^{-1}\frac{\partial\Omega}{\partial x_{k}}}}=v^{-1}(s)
\end{equation}
This relation provides the desired expression for the speed the system
advances along the proposed path. Direct evaluation of the second functional
derivative of the path probability density shows that this is indeed a
maximum. The explicit form of the induced dynamics is%
\begin{equation}
\frac{dx_{i}}{dt}=\sqrt{\frac{\frac{\partial\Omega}{\partial x_{j}}%
g_{kj}^{-1}\frac{\partial\Omega}{\partial x_{k}}}{\frac{dx_{l}}{ds}g_{lm}%
\frac{dx_{m}}{ds}}}\frac{dx_{i}}{ds}
\end{equation}
The probability density itself becomes%
\begin{widetext}
\begin{equation}
P\left[ \rho\right] =\mathcal{N}\exp\left( -\frac{1}{2}\int_{0}^{s_{max}}\left[ 
\sqrt{\frac{dx_{i}}{ds}g_{il}\frac{dx_{l}}{ds}}\sqrt{\frac{\partial\Omega }{%
\partial x_{j}}g_{kj}^{-1}\frac{\partial\Omega}{\partial x_{k}}}-\frac{dx_{i}%
}{ds}\frac{\partial\Omega}{\partial x_{i}}\right] ds\right)   \label{path2}
\end{equation}
\end{widetext}
which, with the time eliminated, gives a method to compare the probabilities
for parametrized paths. Notice that this expression is invariant under a
reparametrization of the reaction coordinate, e.g. $ds\rightarrow u\left(
s^{\prime}\right) ds^{\prime}$, so that it is a purely geometric quantity.
In fact, if we further introduce the gradient force%
\begin{equation}
b_{i}=g_{ij}^{-1}\left( \mathbf{x}\right) \frac{\partial\Omega}{\partial
x_{j}}
\end{equation}
it can be written as 
\begin{equation}
P\left[ \rho\right] =\mathcal{N}\exp\left( -\int_{0}^{s_{max}}\left\vert \frac{d\mathbf{x}}{ds}%
\right\vert \left\vert \mathbf{b}\right\vert \sin\left( \frac {\theta\left( 
\frac{d\mathbf{x}}{ds},\mathbf{b}\right) }{2}\right) ds\right)  
\end{equation}
where $\left\vert \mathbf{b}\right\vert =\sqrt{b_{i}g_{il}b_{j}}$ and where $%
\theta\left( \frac{d\mathbf{x}}{ds},\mathbf{b}\right) $ is the angle between 
$\frac{d\mathbf{x}}{ds}$ and $\mathbf{b}$. This form emphasizes the purely
geometric nature of the optimized path. In particular, the parametrization-invariance means that the result is independent of the value of the upper limit, $s_{max}$, so that any convenient rescaling is allowed.

\section{Applications}

The dynamical theory on which these results are based is applicable to,
e.g., globular proteins in solution which can be modeled at the crudest
level as spherical molecules interacting via a short-ranged effective
potential and subject to Brownian forces due to the solvent. Here, the model
potential of ten Wolde and Frenkel consisting of a hard core and
short-ranged attraction,%
\begin{equation}
V(r)\,=\left\{ 
\begin{array}{c}
\infty,\;\;r<\sigma \\ 
\frac{4\,\epsilon}{\alpha^{2}}\left( \,\left( \frac{1}{(\frac{r}{\sigma }%
)^{2}-1}\right) ^{6}-\,\alpha\,\left( \frac{1}{(\frac{r}{\sigma})^{2}-1}%
\right) ^{3}\right) ,\;\;r\geq\sigma%
\end{array}
\right. \,
\end{equation}
will be used. The energy scale is set by $\epsilon$ and the hard-core radius
is $\sigma$ while the parameter $\alpha$ determines the distance of the
attractive minimum from the hard core:\ the typical value for globular
proteins of $\alpha=50$ will be used. The bulk free energy is approximated
using thermodynamic perturbation theory.

\begin{figure}[tbp]
\includegraphics[angle=-90,scale=0.3]{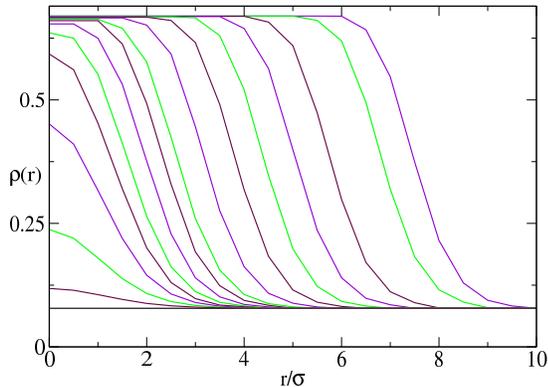} 
\caption{The spherically-symmetric concentration (density) distribution at various points along the classical path, as determined from Eq.(\ref{classical}). The initial state has uniform concentration, $\rho(r) = 0.075$, and the value of the concentration at the origin increases monotonically along the nucleation pathway.}
\label{fig1}
\end{figure}

\begin{figure}[tbp]
\includegraphics[angle=-90,scale=0.3]{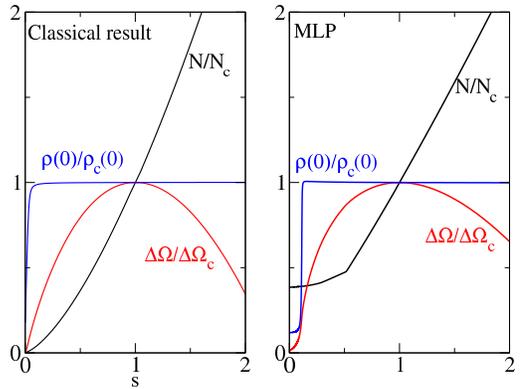} 
\caption{Some aspects of the nucleation pathway as determined using a typical DFT method, left panel, and determined from the dynamical theory, right panel. The figures show the central density, $\rho(r=0)$, excess particle number,$N$, and excess free energy, $\Delta \Omega$, as functions of distance along the nucleation pathway. The left panal, labeled ``Classical'', is the result of the heuristic theory, Eq.(\ref{classical}), whereas the right panel shows the most likely path (MLP), as described in the text. All quantities, including the ordinate, have been scaled by their values at the critical cluster. (Note that despite the differences in the paths, the \emph{critical cluster} is uniquely determined by the free energy and so is the same for both paths.) }%
\label{fig2}
\end{figure}

\begin{figure}[tbp]
\includegraphics[angle=-90,scale=0.3]{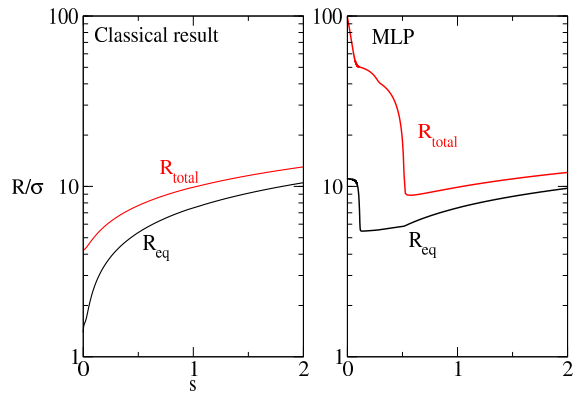} 
\caption{The same as Fig. (\ref{fig2}) but showing the equimolar radius, $R_{\mathrm{eq}}$ and the ``total'' radius, $R_{\mathrm{total}}$, as a function of distance along the nucleation pathway for the path determined from DFT gradient descent, left panel, and from the dynamical theory, right panel.}
\label{fig3}
\end{figure}

In CNT, it is assumed that the material in the interior of a cluster is in
the bulk state:\ e.g., if a liquid phase is being nucleated from a gas then
it is assumed that the density inside the cluster is always that of the bulk
liquid and the cluster grows simply by increasing its radius from zero. More
detailed non-classical results allow for the possibility of the interior
density to vary from the bulk value and the typical behavior, as shown in
Fig. 1, is for the density to begin at that of the background gas and to
increase as the radius increases until eventually reaching that of the bulk
when the cluster is very large. This is associated with the fact that in
contrast to CNT the interface between the cluster and the gas is of finite
extent so that when the cluster is small, all molecules may be considered to
be in an intermediate, interfacial region. Recently, it has been shown that
solving the equations for the MLP yields an unexpected result:\ the cluster
does not begin with a small, localized increase in density but rather it
starts as a spatially extended density fluctuation with the actual
nucleation event occuring within this structure. The initial radius is
infinite and the excess mass is finite so that the density is that of the
vapor. The first part of the process involves the gradual decrease in the
radius with the excess mass remaining nearly constant so that the density
increases slowly. This represents the formation of a density fluctuation
containing excess mass relative to the background. The second stage of the
process is the formation of a nucleus within this region of slightly
enhanced density. The excess mass within the density fluctuation is the
basis for the formation of the cluster which then goes on to grow as in the
classical scenarios.

In order to make a quantitative comparison of these different pathways, calculations of the most likely path for the nucleation of high-concentration droplets in a low-concentration protein solution were performed as described in Ref.\onlinecite{Lutsko_JCP_2012} and using a ``classical'' method which is known to give a good quantitative description of constrained clusters\cite{LutskoGradient}. The determination of the MLP amounts to the solution of Eq.(\ref{gradient}). The classical calculation takes a very similar form to the MLP calculation despite the fact that it was proposed on purely heuristic grounds prior to the MLP method: it involves gradient descent on the free energy surface 
\begin{equation} \label{classical}
\frac{dx_{i}}{ds}=-\frac{1}{\sqrt{\frac{\partial \beta \Delta \Omega }{%
\partial x_{l}}\widetilde{g}_{lk}^{-1}\frac{\partial \beta \Delta \Omega }{%
\partial x_{k}}}}\widetilde{g}_{ij}^{-1}\frac{\partial \beta \Delta \Omega }{%
\partial x_{j}},
\end{equation}%
but with a heuristic metric given by 
\begin{equation}
\widetilde{g}_{ij}\left( \mathbf{x}\right) =\int \frac{\partial \rho \left(
r;\mathbf{x}\right) }{\partial x_{i}}\frac{\partial \rho \left( r;\mathbf{x}%
\right) }{\partial x_{j}}d\mathbf{r.}
\end{equation}%
Note that the first of these equations is equivalent to Eq.(\ref{gradient}), but with distance along the path used as the independent variable rather than ``time'' (since, in the classical theories, there is no dynamics and so no natural concept of time). The similarity between these two calculations is striking and one of the advantages of the dynamical approach is that it shows that gradient descent on the free energy surface does indeed characterize the nucleation pathway provided the correct metric is used. The ``correct'' metric is basically defined in terms of mass differences rather than density differences which can be traced to the fact that the underlying theory is based on fluctuating hydrodynamics in which mass is conserved. In contrast, the heuristic metric was simply guessed based on the criterion of simplicity and a prejudice towards the use of local density as the fundamental variable and, so, no physical meaning can be attached to the paths derived from it. 

In all calculations, spherical symmetry was assumed so that the configuration at any given instant is characterized by the variation of density as a function of distance from the origin. Figure \ref{fig1} shows the density distribution taken from a sequence of points along the classical pathway. The concentration begins as a constant, equal to the concentration of the solution. The formation of a cluster involves a monotonic increase in density near the origin until a density close to that of the bulk high-concentration solution is obtained. Beyond this point, the cluster grows via a monotonic increase in its radius. 

\begin{figure}[tbp]
\includegraphics[angle=-90,scale=0.3]{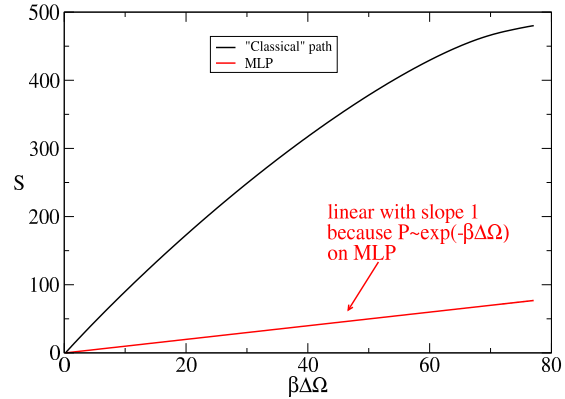} 
\caption{The action for the classical and MLP nucleation pathways shown as a function of excess free energy. Note that in both cases, the path probability density is related to the action by $P \sim e^{-S}$. Also, both paths end at the critical cluster which is the same for all paths. The curve for the MLP is trivial as one knows\cite{Lutsko_JCP_2012} that $S_{MLP} = -\Delta \Omega$.}
\label{fig4}
\end{figure}

For the MLP it is known that the path probability (using either Eq.(\ref%
{path1}) or Eq.(\ref{path2})) is simply given by 
\begin{equation}
P_{MLP}=\mathcal{N}e^{-\beta \Delta \Omega }
\end{equation}%
where $\Delta \Omega $ is the difference in free energy between the
beginning and end points. Numerical evaluation of Eqs.(\ref{path1})-(\ref%
{path2}) confirms this relation to a high degrees of numerical accuracy. 
The ``classical'' model gives the sequence of profiles shown in Fig. 1.
The log of the path probability generated by this ansatz is shown in Figure
2. The energy barrier in these calculations is $\beta \Delta \Omega \sim 88$
so that $P_{MLP}\simeq \mathcal{N}e^{-88}$. In contrast, the probability of
the heuristic path is $P_{classical}\simeq \mathcal{N}e^{-480}$ so that it
is as if the energy barrier were $480k_{B}T$ rather than $88k_{B}T$. The
relative probability is 
\begin{equation}
P_{classical}=e^{-398}P_{MLP}
\end{equation}%
so that the heuristic path is very unlikely compared to the MLP. This simple
comparison illustrates the fact that the path probability is quite sensitive
to the path and is not determined solely by the free energy barrier.

\begin{figure}[tbp]
\includegraphics[angle=-90,scale=0.3]{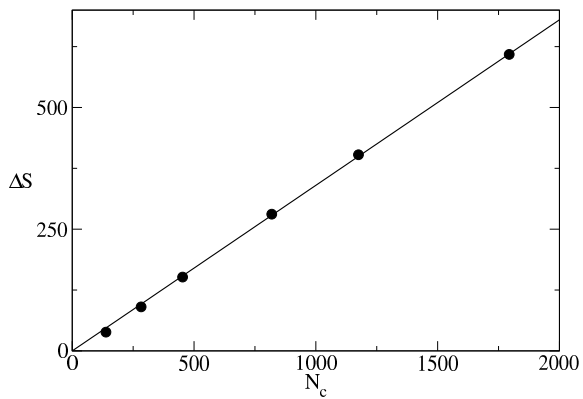} 
\caption{The difference in action between the classical path and the MLP, $\Delta S = S_{classical}-S_{MLP}$,  for different values of supersaturation corresponding to different critical cluster sizes. The relative path probability for the two paths is $P_{classical}/P_{MLP} = e^{-\Delta S}$. A linear fit to the calculated points gives $\Delta S \approx 0.34 N_c$. }
\label{fig5}
\end{figure}
Finally, the variation of the relative path probabilities as a function of the supersaturation is illustrated in Fig. \ref{fig5} which shows the difference in the action as a function of size of the critical cluster. At least for the range of cluster-sizes shown in the figure, there is an almost linear relation which implies the relation $P_{classical} \approx e^{-N_{critical}/3}P_{MLP}$.
\section{Conclusions}

Based on the recently proposed dynamical approach to nucleation, a method
has been derived for determining the maximum probability of observing any
given nucleation pathway. The method is applicable to all pathways without
regard to how they are constructed. This allows one to then ask whether
heuristically-derived pathways are reasonable approximations to the most
likely pathway. The formalism has been illustrated for the problem of the
nucleation of high-concentration droplets from a low-concentration solution
of globular proteins. It was shown that the probability for a "classical"
nucleation pathway, which begins with a small, localized  cluster, compared
to that of the most likely path, which invokes a two-step mechanism whereby
nucleation begins with a spatially-extended, small-amplitude density
fluctuation, scales like $e^{-N_{critical}/3}$  so that for all but the smallest clusters, it is extremely small. 
This comparison serves to show that classical paths, even though intuitively appealing, can have very low probability of occurrence. Contrary to the expectation that all that matters for nucleation is the free energy barrier, these results show that there is a dramatic quantitative difference in the likelihood of observing the classical scenario compared to  the non-classical one. Of course, only one family of ``classical'' pathways has been investigated here and there are many alternatives such as those found in Refs. \onlinecite{talanquer:5190, Corti_PRL, Ghosh, EvansArcherNucleation}. It is possible that some of these compare more favorably with the non-classical Most Likely Path, but this can only be resolved by direct calculations. 

\begin{acknowledgments}
This work was partially supported in part by the European Space Agency under contract number ESA AO-2004-070 and by FNRS Belgium under contract C-Net NR/FVH 972.
\end{acknowledgments}

\end{document}